# Quantifying Divergence in Inter-LLM Communication Through API Retrieval and Ranking

**Eyhab Al-Masri**

School of Engineering and Technology
University of Washington (Tacoma)
ealmasri@uw.edu

## Abstract

Large language models (LLMs) increasingly operate as autonomous agents that reason over external APIs to perform complex tasks. However, their reliability and agreement remain poorly characterized. We present a unified benchmarking framework to quantify inter-LLM divergence—the extent to which models differ in API discovery and ranking under identical tasks. Across 15 canonical API domains and 5 major model families, we measure pairwise and group-level agreement using set-, rank-, and consensus-based metrics: Average Overlap, Jaccard, Rank-Biased Overlap, Kendall's τ/W, and Cronbach's α. Results show moderate overall alignment (AO ≈ 0.50, τ ≈ 0.45) but strong domain dependence: structured tasks (Weather, Speech-to-Text) are stable, while open-ended ones (Sentiment Analysis) diverge sharply. Volatility and consensus analyses reveal that coherence clusters around data-bound domains and degrades for abstract reasoning. These insights enable reliability-aware orchestration in multi-agent systems, where consensus weighting can improve coordination among heterogeneous LLMs. Beyond performance benchmarking, our results reveal systematic failure modes in multi-agent LLM coordination, where apparent agreement masks instability in action-relevant rankings. This hidden divergence poses a pre-deployment safety risk, motivating diagnostic benchmarks for early detection.

## Introduction

Large Language Models (LLMs) are increasingly deployed as autonomous agents that plan, reason, and act within complex environments through external tool invocation (Yao et al., 2023; Schick et al., 2023). This agentic paradigm enables tasks requiring real-time data access, structured computation, and interaction with digital systems, shifting the LLM's role from text generation to action orchestration through tool and API calls (Tzachristas et al., 2023; Li et al., 2023). Recent orchestration frameworks now connect multiple LLMs that communicate, critique, and coordinate, forming the foun dation of emerging multi-agent reasoning systems (Li et al., 2024a).

The enhanced utility of these agents stems from Tool Learning, where LLMs extend beyond pre-trained knowledge by dynamically discovering and invoking external APIs (Schick et al., 2023; Xu et al., 2025). This enables models to transform natural-language goals into executable API workflows involving discovery, parameter generation, and multi-step composition (Huang et al., 2024; Morais et al., 2025). However, mapping natural instructions to structured API calls remains fragile—highly sensitive to prompting strategies and configuration choices (Sheng et al., 2024).

Within multi-agent settings, this fragility manifests as inter-LLM divergence: when presented with identical tasks and toolsets, different models frequently disagree on which APIs are relevant and how they should be prioritized (Al-Masri et al., 2025). This phenomenon—inter-LLM API ranking divergence—represents inconsistency in the reasoning process that precedes any downstream execution. Such disagreement undermines coordinated planning, reproducibility, and verifiable action among autonomous agents.

We hypothesize that this divergence is structured rather than random—that is, LLMs exhibit domain-dependent reliability patterns, converging on well-defined, data-bound tasks and diverging on creative or semantically open ones. To test this, we systematically benchmark the extent and structure of inter-LLM divergence in API retrieval and ranking across multiple model families and task domains.

Understanding how and when LLMs diverge provides a basis for designing coordination strategies among agents that rely on shared tool-use and retrieval pipelines. Quantifying divergence clarifies the limits of reproducibility and stability in reasoning-driven orchestration. This study formalizes divergence as an observable reliability dimension within multi-agent reasoning pipelines. Specifically, this paper makes the following contributions:

- **Quantitative Benchmarking Framework:** We develop a reproducible pipeline using the AutoGen multi-agent library (Wu et al., 2024) to evaluate inter-LLM divergence in API retrieval under uniform conditions.
- **Comprehensive Cross-Model Evaluation:** Using fifteen canonical API-discovery tasks and five major LLM

families (ChatGPT, Claude, Gemini, DeepSeek, and Mistral), we generate 750 ranked lists and assess consistency through multi-level agreement metrics.
- **Reliability and Consensus Metrics:** We integrate set-based, rank-based, and group reliability measures—Average Overlap, Jaccard, Rank-Biased Overlap, Kendall's τ/W, and Cronbach's α—to quantify pairwise and collective coherence.
- **Divergence Structure and Latent Analysis:** We extend evaluation with ANOVA and Kruskal–Wallis significance testing, and volatility–agreement regression to validate inter-model patterns.

From a safety perspective, inter-LLM divergence is not merely a performance artifact but a potential precursor to deceptive or unintended coordination. When multiple agents appear to agree at the surface level (e.g., shared API retrieval) yet diverge sharply in internal prioritization, this creates conditions under which downstream behavior may become brittle, non-reproducible, or strategically misaligned. Recent concerns in AI safety highlight that such discrepancies can enable hidden failure modes that evade standard evaluation yet manifest during deployment. By making these divergences measurable, this work contributes a diagnostic foundation for identifying coordination risks before autonomous agents are deployed at scale.

This study introduces a unified empirical basis for evaluating reliability and coherence in LLM-based API reasoning. By quantifying how model agreement varies across task domains and ranking depths, it establishes a benchmark for analyzing reasoning stability in multi-agent systems. The framework bridges prior work on tool learning and service discovery with emerging research on agent consistency—providing a reproducible foundation for examining when, and why, large language models converge or diverge in their interpretation of shared tasks.

## Related Work

Research on the automated discovery of network-accessible functionalities originated in the web service era, where crawlers indexed and benchmarked SOAP-based services for quality and structure (Benatallah et al., 2005; Al-Masri et al., 2010). This foundational work established the need for large-scale observation and evaluation of distributed services. The transition to lightweight, RESTful architectures marked a major shift, extending discovery to developer-focused web APIs exposed over HTTP (Richardson et al., 2013). In this phase, web search engines became the de facto mechanism for locating APIs, relying on keyword matching of documentation—a process later enhanced by structured specifications such as OpenAPI (Ponelat et al., 2022).

Building on these advances, researchers have increasingly investigated how large language models (LLMs) enhance API discovery, selection, and composition, moving from keyword-based retrieval toward semantic and reasoning-driven automation (Bianchini et al., 2025; Pesl et al., 2023). LLMs interpret natural-language goals and infer suitable APIs, acting as intelligent intermediaries between intent and execution. To address sparsity in service networks, several studies apply LLM-based semantic enrichment (Peng et al., 2025) or integrate graph neural networks for improved alignment between textual and structural data (Feng et al., 2025), though such methods remain domain-limited and data-dependent.

Parallel work explores LLM-based service composition and validation, where models autonomously generate, test, and execute API workflows. Examples include decision-tree reasoning for API selection (Zhang et al., 2024), conversational DevOps automation through ChatOps4Msa (Wang et al., 2024), automated logical testing with multi-agent LLMs (Zhang et al., 2025), and API-first specification generation (Chauhan et al., 2025). Additional studies extend LLMs to API testing and specification synthesis, improving automation and coverage in service-oriented architectures (Altin et al., 2025; Zheng et al., 2024; Smardas et al., 2025; Deng et al., 2025).

Despite this progress, prior research has largely focused on what LLMs can produce—discovering, composing, or validating APIs—without assessing how consistently they reason or converge across model families (Al-Masri et al., 2025). This gap becomes critical as LLMs evolve into autonomous agents operating across domains such as IoT, edge computing, and machine learning (Aiello et al., 2023; Aiello, 2025). In multi-agent contexts, the reliability of API-driven reasoning determines whether agents can coordinate actions, maintain shared world models, and yield stable collective outcomes. This study addresses that gap by systematically benchmarking inter-LLM divergence in API ranking and selection, providing the first empirical assessment of reliability and agreement in LLM-based multi-agent reasoning.

## Inter-LLM Divergence Framework

To quantify reasoning consistency across large language models (LLMs), we developed an automated benchmarking framework using the AutoGen multi-agent orchestration library (Wu et al., 2024). The system enables concurrent querying and analysis of multiple model APIs under uniform conditions, ensuring that observed differences reflect model reasoning rather than prompt variation.

The framework operates through six sequential modules:
(1) **Query Generation** – Encodes canonical task templates and dispatches them with fixed temperature and reasoning depth.
(2) **Model Execution** – Runs each LLM agent on identical API-ranking prompts and records structured JSON outputs.

(3) **Validation** – Normalizes field names, removes duplicates, and ensures schema compliance.
(4) **Scoring** – Extracts rank order, relevance percentage, and justification text.
(5) **Summarization** – Consolidates outputs into unified tables for statistical and comparative analysis.
(6) **Agreement Analysis** – Computes similarity, reliability, and consensus metrics across model pairs and groups.

All analyses were implemented in Python 3.11 using NumPy and Pandas, ensuring replicability of all metric computations. This modular design enables reproducible, large-scale comparison across LLM families while preserving semantic interpretability. Unlike prior work that evaluates models in isolation, our approach emphasizes cross-model relational reliability—how consistently different LLMs identify and prioritize the same APIs under identical conditions. To assess statistical significance across metrics and model families, we applied one-way ANOVA, Kruskal–Wallis, and Levene's tests; no post-hoc comparisons were required, as no significant group-level differences were observed.

## Dataset and Prompt Construction

The benchmark encompasses fifteen canonical API-discovery tasks drawn from the Postman 2025 State of the API Report (Postman, 2025). These tasks span diverse domains such as image generation, speech-to-text, event discovery, weather forecasting, and financial data access. Each task was encoded as a canonical query template defining both format and content expectations. For instance:

*"Return exactly 10 hosted web APIs for discovering upcoming events. Each must accept search parameters such as location, date, or category and return JSON with event name, date, venue, location, and description."*

Each model received the same prompt and produced a structured JSON response containing ten ranked APIs per query. Each entry included nineteen attributes—covering structural metadata (e.g., endpoints, authentication, rate limits), functional metadata (e.g., pricing, response schema), and semantic reasoning fields (e.g., rationale, evidence, limitations).

In total, the dataset comprises 14,250 atomic data points (15 queries × 5 LLMs × 10 APIs × 19 attributes), forming a unified basis for large-scale quantitative analysis of reasoning consistency. All benchmark scripts, analysis notebooks, and aggregated datasets are publicly available on https://github.com/aeris-lab/llmrank to support replication and further research. The processed dataset was then used for quantitative and semantic analysis, enabling consistent comparison across LLM families and task domains.

## Quantitative and Semantic Data Representation

All model outputs were flattened and merged into a comparative analytical dataset using Python-based preprocessing in Jupyter. Each record was annotated by model family, task domain, and rank position, enabling pairwise and group-level comparisons across LLMs. The evaluated models—OpenAI ChatGPT-5, Anthropic Claude 4.5 Sonnet, Google Gemini 2.5 Flash, DeepSeek, and Mistral Large 2—were selected to represent both proprietary and open-source model families widely used in multi-agent systems.

We computed a suite of complementary metrics capturing different layers of inter-model consistency:

**Set-Based Similarity**

o Average Overlap (AO): Measures the mean cumulative agreement depth between two ranked lists *A* and *B* up to position $k$. At each rank depth $d$, the overlap between the top-$d$ items of both lists is computed and averaged across all depths. Higher AO values indicate stronger overlap throughout the ranking, not just in the top results. (Webber et al., 2010).

$$\text{AO}(A, B) = \frac{1}{k}\sum_{d=1}^{k} \frac{|A_{1:d} \cap B_{1:d}|}{d} \quad (1)$$

where:
- A,B - ranked lists of APIs produced by two LLMs;
- $A_{1:d}, B_{1:d}$ - subsets of A and B containing top-d APIs;
- k - total ranking depth considered (here, k=10);
- $|A_{1:d} \cap B_{1:d}|$ - count overlapping APIs up to depth d.

o Jaccard Similarity (J): Measures the proportion of shared APIs between two ranked lists *A* and *B*. It computes the ratio of the intersection to the union of the sets, representing how frequently both models identify the same APIs as relevant, regardless of order (Jaccard, 1901; Costa, 2021). Higher Jaccard values indicate greater retrieval overlap and shared coverage of the API space.

$$J(A, B) = \frac{|A \cap B|}{|A \cup B|} \quad (2)$$

where *A, B* represent sets of APIs retrieved by LLMs for same task; |A∩B| represents the number of APIs common to both models; |A∪B| represents the total number of distinct APIs retrieved; and J(A, B) represents a range from 0 (no overlap) to 1 (full overlap).

**Rank-Sensitive Similarity**

o Rank-Biased Overlap: Extends the Average Overlap measure by incorporating a top-weighting parameter $p$ that emphasizes agreement near the top of ranked lists. (Webber et al., 2010; Corsi et al., 2024). The score decays geometrically with depth, giving higher importance to early ranks—an essential property when top APIs drive execution success. RBO captures both the extent of overlap and how early in the ranking that overlap occurs.

$$\text{RBO}(A, B, p) = (1-p)\sum_{d=1}^{k} p^{d-1} \frac{|A_{1:d} \cap B_{1:d}|}{d} \quad (3)$$

where A, B represent ranked lists of APIs from two LLMs; $A_{1:d}, B_{1:d}$ prefixes of the lists up to depth d; k represents the maximum rank depth considered (here,

10); p represents the persistence or top-weighting parameter (0 < p < 1); $|A_{1:d} \cap B_{1:d}|$ represent the number of shared items among the top-d APIs; and RBO(A, B, p) ranges from 0 (no overlap) to 1 (identical ranked lists).

o Kendall's τ (Tau) Rank Correlation: Measures the ordinal agreement between two ranked lists by comparing the number of concordant and discordant API pairs (Kendall, 1938; McLeod et al., 2005). A concordant pair preserves the same order across both lists, while a discordant pair inverts it. Higher τ values indicate stronger consistency in ranking structure, whereas negative values signal rank inversions. Unlike RBO, which emphasizes early ranks, τ treats all positions equally and captures global rank correlation.

$$\tau = \frac{C - D}{0.5n(n-1)} \quad (4)$$

where C represents the number of concordant pairs of APIs (ordered identically in both lists); D represents the number of discordant pairs (order reversed between lists); n represents the total number of ranked items (here, n=10); 0.5 n(n−1) represents the total number of possible pairwise comparisons; and τ ranges from −1 (complete inversion) to +1 (perfect agreement), with 0 representing random ordering.

**Group Reliability and Internal Consistency**

o Kendall's W (Coefficient of Concordance): Extends Kendall's τ to multiple raters (here, LLMs) and measures the overall strength of agreement among all models. (Kendall et al., 1939; Abdi et al., 2007). It quantifies how consistently $m$ models rank $n$ items, producing a single group-level reliability score. $W$=1 indicates perfect consensus among models, while $W$=0 represents complete independence of rankings.

$$W = \frac{12 \sum_{i=1}^{n}(R_i - \bar{R})^2}{m^2(n^3 - n)} \quad (5)$$

where $R_i$ represents the sum of ranks assigned to item $i$ across all $m$ models; $\bar{R}$ represents the mean of all $R_i$ values; $m$ represents the number of raters or models (here, 5 LLMs); $n$ represents the number of ranked APIs per task (10); and $W$ is the coefficient of concordance representing group-level ranking agreement.

o Cronbach's α (Alpha): Measures the internal consistency or reliability of a group of models treated as raters (Cronbach, 1951; Cho et al., 2015). It assesses how closely related their rankings are as a group—higher α values indicate that models behave coherently when evaluating API relevance. In this study, α complements Kendall's W by providing a variance-based measure of inter-model cohesion.

$$\alpha = \frac{k}{k-1}\left(1 - \frac{\sum_{i=1}^{k} \sigma_i^2}{\sigma_t^2}\right) \quad (6)$$

where $k$ represents the number of items or models (here, $k$=5); $\sigma_i^2$ represents the variance of each model's assigned relevance scores across APIs; $\sigma_t^2$ represents the total variance of the combined scores across all models; and $\alpha$ is an internal consistency coefficient (0 ≤ α ≤ 1). Values above 0.8 typically indicate strong reliability, while lower values suggest inconsistency in model reasoning.

**Uncertainty and Consensus Metrics**

o Volatility Score (V($a_j$)): Quantifies the dispersion of ranks assigned to the same API $a_j$ across all $m$ models, following the approach for measuring model uncertainty in ranking (Cohen et al., 2021). A high V($a_j$) indicates significant disagreement and high uncertainty among models about an API's true rank/importance, while a low V($a_j$) suggests stable confidence in its ranking.

$$V(a_j) = \text{Var}\left(\text{rank}_{a_j}(m)\right) \quad (7)$$

where $a_j$ represents the j$^{th}$ API; rank$_{aj}$(m) is the rank assigned by model m; Var(·) captures the spread of assigned ranks; m is the number of models; and V($a_j$) is rank variance. To aggregate these uncertainties at the query level, we define the Average Ranking Volatility (ARV) as the mean of the Volatility Scores of all unique APIs ($A_Q$) for that query, providing an overall measure of rank stability for the search domain:

$$\text{ARV}(Q) = \frac{1}{|A_Q|} \sum_{a_j \in A_Q} V(a_j) \quad (8)$$

o Kemeny–Young Consensus Distance ($D_K$): Measures the degree of global disagreement among all model rankings (Kemeny, 1959; Young et al., 1978; Azzini et al., 2020). $D_K$ is the normalized distance between the observed rankings and the consensus ranking that minimizes total pairwise discord. It is practically computed as the normalized difference between the average ranks of all API pairs. Lower $D_K$ values indicate strong cross-model consensus, while higher values reflect fragmented or contradictory ranking preferences.

$$D_K = 1 - \frac{1}{N_{pairs}} \sum_{i<j} \left(1 - \frac{|r_i - r_j|}{n-1}\right) \quad (9)$$

where $r_i$, $r_j$ represents the average ranks of APIs i and j across all models; n number of ranked APIs (here, n=10); $N_{pairs}$=(0.5(n(n−1))) represents the total number of API pairs compared; $|r_i - r_j|$ represents the absolute rank difference between APIs i and j; $D_K$ represents the consensus distance ranging from 0 (perfect agreement) to 1 (maximal disagreement). In practice, $D_K$ can be approximated as 1−$\bar{\tau}$, where $\bar{\tau}$ is the mean pairwise Kendall correlation.

We next evaluate inter-LLM divergence across fifteen API-discovery tasks and five model families.

# Results and Evaluation

We evaluated inter-LLM API ranking divergence across 15 canonical tasks and five model families (ChatGPT-5, Claude 4.5 Sonnet, Gemini 2.5 Flash, DeepSeek, and Mistral-Large 2). Each model generated ten ranked APIs per task, producing 750 ranked lists analyzed through our benchmarking pipeline. Pairwise agreement was measured using Jaccard, RBO, and Kendall's τ; group reliability via Kendall's W and Cronbach's α; and reasoning uncertainty through Volatility–Agreement regression and statistical tests (ANOVA, Kruskal–Wallis, Levene).

## Pairwise Agreement and Divergence Patterns

Table 1 summarizes the quantitative agreement across all large language models (LLMs) and evaluation metrics—Average Overlap (AO), Jaccard Similarity, Rank-Biased Overlap (RBO, p = 0.9), and Kendall's τ—computed over fifteen API discovery tasks. Figure 1 visualizes the pairwise similarity matrix complementing Table 1, while Figure 2 aggregates them into a cumulative view of similarity strength by model pair. Collectively, these reveal a structured but asymmetric reliability landscape in inter-LLM reasoning.

Across metrics, agreement remains moderate overall (mean AO ≈ 0.50, Jaccard ≈ 0.36, RBO ≈ 0.33, τ ≈ 0.45), confirming that LLMs partially converge on retrieved APIs but diverge in rank prioritization. The highest overall stability occurs between Claude–DeepSeek (AO = 0.58, τ = 0.65), followed by Claude–Mistral (AO = 0.56, τ = 0.62) and DeepSeek–Mistral (AO = 0.53, τ = 0.62). Claude–Gemini also shows high retrieval overlap (AO = 0.56) but lower rank-order correlation (τ = 0.41), indicating agreement in discovery yet divergence in prioritization. In these pairs, both overlap and ordering are strong, implying aligned reasoning heuristics and similar criteria for API relevance. The τ values above 0.60 indicate that over 60% of pairwise API orderings are concordant—approaching the inter-annotator reliability typically observed in information-retrieval benchmarks.

A one-way ANOVA, Kruskal–Wallis, and Levene's tests confirmed no statistically significant differences across model pairs for any metric (all p > 0.4). This supports that inter-LLM variation arises from structured reasoning behavior rather than random noise, forming a continuous reliability gradient across models.

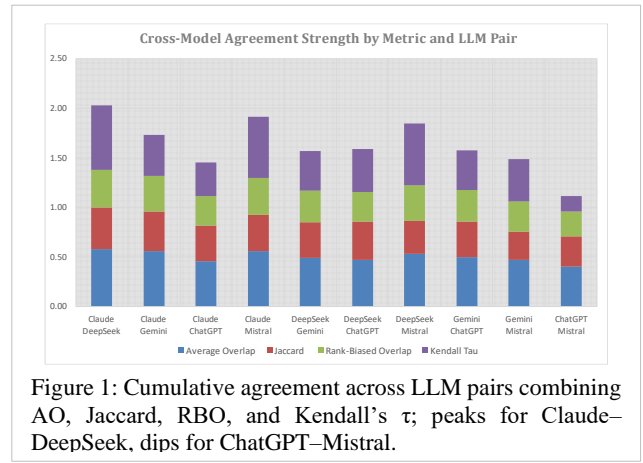

Figure 1: Cumulative agreement across LLM pairs combining AO, Jaccard, RBO, and Kendall's τ; peaks for Claude–DeepSeek, dips for ChatGPT–Mistral.

In contrast, Gemini–Mistral (τ = 0.43) and ChatGPT–Mistral (τ = 0.15) show the weakest alignments. The latter indicates near-random rank correlation—despite some overlap (AO = 0.40, Jaccard = 0.30)—as both differ sharply in how they prioritize APIs. Such divergence likely stems from differences in fine-tuning data and optimization goals rather than task framing. Metric-specific patterns further explain this heterogeneity.

- **Average Overlap (0.40–0.58)** captures shared discovery space: models tend to identify roughly half of the same APIs, consistent with partial semantic convergence.
- **Jaccard Similarity (0.29–0.42)** normalizes for set size, revealing that beyond this shared core, each model still introduces a distinct portion of the API space.
- **Rank-Biased Overlap (0.26–0.38)** drops significantly relative to AO, confirming that agreement declines for top-ranked items—the decisions most critical for execution in multi-agent systems.
- **Kendall's τ (0.15–0.65)** offers the clearest ordering signal: high-τ pairs (> 0.6) show coherent ranking logic, while low-τ ones (< 0.3) reflect stochastic prioritization.

The combined metrics reveal a three-tier reliability hierarchy based on joint thresholds of AO and Kendall's τ.

- **High-stability pairs:** Claude–DeepSeek, Claude–Mistral, and DeepSeek–Mistral—exhibit strong convergence in both retrieval and ranking (AO ≥ 0.53, τ ≥ 0.60).
- **Moderate-stability pairs:** Claude–Gemini, DeepSeek–Gemini, Gemini–ChatGPT, DeepSeek–ChatGPT, and

| Metric | Claude DeepSeek | Claude Gemini | Claude ChatGPT | Claude Mistral | DeepSeek Gemini | DeepSeek ChatGPT | DeepSeek Mistral | Gemini ChatGPT | Gemini Mistral | ChatGPT Mistral |
|---|---|---|---|---|---|---|---|---|---|---|
| Average Overlap | 0.58 | 0.56 | 0.46 | 0.56 | 0.49 | 0.47 | 0.53 | 0.50 | 0.47 | 0.40 |
| Jaccard | 0.42 | 0.40 | 0.36 | 0.36 | 0.36 | 0.39 | 0.34 | 0.35 | 0.29 | 0.30 |
| Rank-Biased Overlap | 0.38 | 0.36 | 0.29 | 0.37 | 0.32 | 0.30 | 0.36 | 0.32 | 0.31 | 0.26 |
| Kendall Tau | 0.65 | 0.41 | 0.34 | 0.62 | 0.39 | 0.43 | 0.62 | 0.40 | 0.43 | 0.15 |

Table 1. Mean pairwise agreement scores across LLMs and metrics.

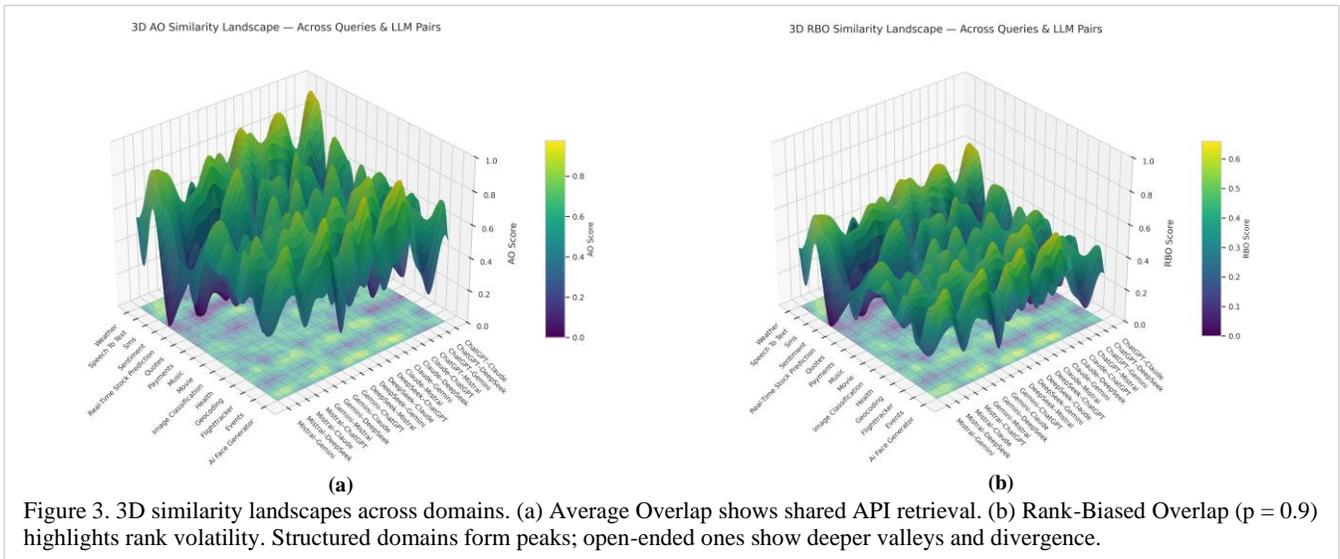

Figure 3. 3D similarity landscapes across domains. (a) Average Overlap shows shared API retrieval. (b) Rank-Biased Overlap (p = 0.9) highlights rank volatility. Structured domains form peaks; open-ended ones show deeper valleys and divergence.

Gemini–Mistral—show consistent overlap (AO ≈ 0.47–0.56) but weaker rank alignment (τ ≈ 0.40–0.43).
- **Lower-stability pairs**, Claude–ChatGPT and ChatGPT–Mistral, fall below these thresholds (AO < 0.46, τ < 0.35), with both retrieval and order coherence breaking down.

This pattern shows a clustered reliability structure, with Claude bridging proprietary models (ChatGPT, Gemini) and open-source ones (DeepSeek, Mistral). Figure 1 underscores this contrast—Claude pairs dominate the upper range, while ChatGPT–Mistral anchors the lowest tier.

## Domain-Level Reliability Trends

Following pairwise comparisons, Figure 2 extends the evaluation to the domain level, showing how reasoning consistency varies across fifteen canonical task categories. Each stacked bar aggregates the four normalized metrics (AO, Jaccard, RBO, τ) per domain.

Structured, well-defined domains—such as Geocoding, SMS, Speech-to-Text, and Weather—demonstrate the highest cumulative reliability (total ≈ 2.1–2.4). These tasks provide clear functional constraints and consistent documentation patterns, enabling LLMs to converge on similar API endpoints and ranking orders. For example, weather and geocoding APIs share predictable parameters (e.g., location, units, coordinates), minimizing interpretive variability.

Conversely, semantically open or creative domains—AI Face Generator and Sentiment Analysis—show the weakest overall agreement (total ≤ 1.0). In these tasks, LLMs diverge widely both in which APIs they retrieve and in how they justify their selections. The extremely low RBO and τ values (≤ 0.35) confirm that top-ranked APIs vary substantially, reflecting unstable reasoning heuristics when explicit task constraints are absent.

Low- to moderate-stability domains—such as Health, Payments, and Flight Tracker—exhibit partial convergence (AO ≈ 0.31–0.41) but weak rank consistency (τ ≈ 0.15–0.46). Models tend to agree on the broad service category yet differ in ranking emphasis—for instance, prioritizing free versus commercial APIs or structured versus unstructured data sources.

Intermediate domains—including Music, Movie, and Events—show higher stability (total ≈ 2.1–2.25) than the low group but remain slightly below infrastructure tasks. These categories benefit from standardized media formats and widely adopted API conventions, fostering more coherent reasoning across models.

The average bar at the top of Figure 2 provides a concise meta-summary, indicating that while the average collective agreement across all tasks is approximately 2.0, high-stability domains can score significantly higher (up to ≈ 2.4). This aligns with the moderate cross-pair correlations observed earlier and underscores that inter-LLM reasoning variability persists even in standardized task templates. These systematic variations suggest that model coordination should be domain-adaptive—structured domains enable consistent collaboration, while abstract domains may require consensus filtering or selective routing among agents.

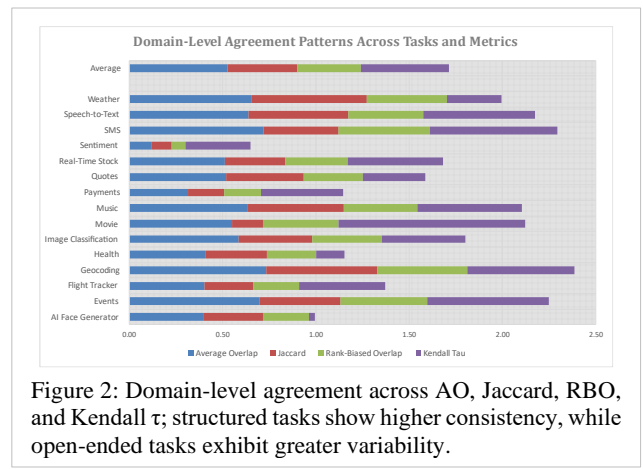

Figure 2: Domain-level agreement across AO, Jaccard, RBO, and Kendall τ; structured tasks show higher consistency, while open-ended tasks exhibit greater variability.

## Cross-Metric Synthesis and Observations

Collectively, the quantitative results from Table 1 and Figures 1–2 establish that inter-LLM divergence is systematic, domain-sensitive, and cluster-dependent. Proprietary models—ChatGPT, Claude, and Gemini—show moderate internal alignment, while open-source counterparts—DeepSeek and Mistral—exhibit broader variability. Cross-model clusters such as Claude–DeepSeek–ChatGPT achieve the highest overall stability ($\tau \approx 0.62$–$0.65$), whereas ChatGPT–Mistral ($\tau = 0.15$) and Gemini–Mistral ($\tau = 0.43$) illustrate divergent reasoning paths. The domain-level results further confirm that reasoning coherence improves when external structure constrains interpretation and declines when tasks require subjective or creative inference.

From a systems perspective, these findings indicate that reliable multi-agent orchestration cannot depend on naïve majority voting or pairwise consensus alone. Instead, weighted consensus mechanisms that emphasize high-reliability clusters (e.g., Claude–DeepSeek–ChatGPT) and down-weight volatile domains can enhance coordinated decision-making. This evidence-based mapping of reasoning coherence offers a quantitative foundation for consensus calibration and trust-aware orchestration in LLM-driven multi-agent frameworks.

## Similarity Landscape Analysis

To assess cross-model reliability, we visualize 3D similarity landscapes using Average Overlap (AO) and Rank-Biased Overlap (RBO) (Figures 3a–b). These topographical surfaces show agreement peaks and divergence valleys across task domains. The AO landscape highlights structured tasks (Weather, Speech-to-Text, Flight Tracker) with broad agreement and creative domains (AI Face Generator, Sentiment Analysis) with lower overlap. Its smooth surface suggests models share a common retrieval base despite differing rank priorities.

The RBO landscape (Figure 3b) shows sharper peaks and troughs, reflecting stronger variation in rank-order stability. High RBO regions coincide with structured domains, while deeper troughs mark areas where LLMs disagree on API ordering despite partial overlap. Notably, pairs involving Mistral and Gemini form consistently lower RBO regions, confirming that rank-sensitive disagreement is concentrated among models with greater exploratory or generative variance.

Together, the two surfaces illustrate that inter-LLM reliability is both domain-dependent and rank-sensitive: structured tasks promote consistent API retrieval, whereas subjective or open-ended domains amplify ranking divergence across models.

## Rank-Depth Evolution of Model Agreement

To complement the 3D landscapes, Figure 4 illustrates how Average Overlap (AO) evolves across rank depths for all LLM pairs.

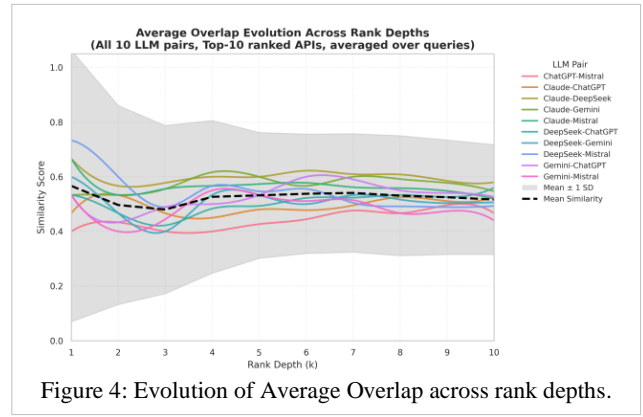

Figure 4: Evolution of Average Overlap across rank depths.

Similarity peaks in the top-rank region ($k \leq 3$), where models most strongly agree on the most salient APIs, and gradually stabilizes near 0.5 as depth increases. The narrowing ±1 SD band indicates convergence in lower ranks, suggesting that once peripheral APIs are reached, outputs become more uniform despite early-rank disagreement.

Claude–DeepSeek maintain the highest overlap (AO = 0.58), and Claude-Gemini shows high overlap (AO=0.56), implying consistent retrieval preferences in these pairs. Conversely, ChatGPT–Mistral shows the greatest volatility. This trend reinforces the landscape findings: strong early-rank alignment in structured tasks (e.g., Weather, Speech-to-Text) and increasing divergence in creative or loosely defined domains.

The findings indicate that inter-LLM variability is primarily driven by a split between retrieval agreement and rank dependency. While models share a common pool of retrieved items, their disagreement in prioritizing the top ranks explains the majority of the observed divergence.

## Stability and Consensus Analysis

To assess how consistently large language models (LLMs) prioritize APIs across domains, we analyzed ranking volatility and global consensus (Figures 5-6).

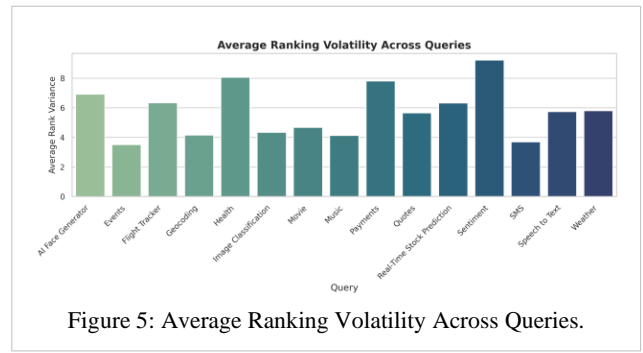

Figure 5: Average Ranking Volatility Across Queries.

Ranking volatility (Figure 5) quantifies how much API positions fluctuate across models for each domain. Structured tasks such as Weather, Speech-to-Text, and Events exhibit the lowest variance ($\approx$ 3–4), reflecting stable reasoning and consistent prioritization. In contrast, open-ended domains such as Sentiment Analysis, Health, and AI Face Generator show the highest volatility (> 7), revealing greater interpretive diversity and model-specific sensitivity to prompt formulation.

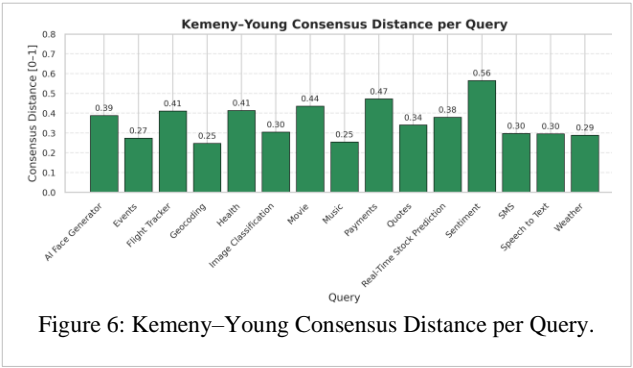

Figure 6: Kemeny–Young Consensus Distance per Query.

Volatility and consensus distance are inversely related, confirming that greater dispersion directly translates into weaker global agreement. The Kemeny–Young consensus distance (Figure 6) aggregates all pairwise rankings into a single consensus order to quantify overall alignment. Lower distances (< 0.6) in tasks such as Geocoding and Speech-to-Text reflect cohesive ranking behavior, whereas higher values (> 0.9) for Sentiment Analysis and Health indicate fragmented reasoning and limited agreement among LLMs.

In combination, these analyses confirm that cross-model stability is strongly domain-dependent—LLMs converge in structured, data-driven contexts but diverge sharply in tasks requiring subjective interpretation or semantic abstraction. Taken as a whole, the findings reveal a structured reliability landscape, where agreement patterns mirror the broader trends discussed later on reasoning coherence and reliability in multi-agent systems.

## Discussion

The analyses provide a comprehensive view of how LLMs diverge and converge when reasoning over identical API-discovery tasks. Across all metrics, consistent patterns clarify reliability limits. The most critical finding is the significant rank instability: while models show moderate retrieval overlap (Mean AO ≈ 0.50), this drops sharply for top-ranked items (Mean RBO ≈ 0.33). This gap proves that models agree on what APIs are relevant but strongly disagree on which should be executed first. The results point to an emerging taxonomy of agreement shaped jointly by domain structure and each model's internal ranking logic.

At the global level, the pairwise heatmap (Figure 1) and summary table (Table 1) show moderate overall agreement, with mean AO clustering around 0.50 and Jaccard around 0.36. These values indicate partial but not complete alignment: LLMs select overlapping API sets yet differ in fine-grained ordering. The reliability landscape is defined by clear clusters: the Claude-DeepSeek pair achieves the highest rank correlation (AO ≈ 0.60, τ ≈ 0.65), forming a high-trust cluster[4]. Conversely, the ChatGPT-Mistral pair exhibits near-random ranking (AO ≈ 0.40, τ ≈ 0.15), representing a persistent structural divergence[5].

A one-way ANOVA and Kruskal–Wallis test found no statistically significant differences across model pairs for any metric (all p > 0.4), confirming that observed variations in similarity are structural rather than random. This supports the interpretation that inter-LLM divergence follows a continuous reliability gradient rather than discrete performance clusters.

Figure 1 further confirms that correlation-based measures (τ) are generally higher than set-based ones (Jaccard), indicating that inter-model disagreement is largely ordinal rather than categorical. While pairwise patterns outline aggregate reliability, the domain-level view reveals how structural constraints drive convergence.

Figures 1 and 2 expose strong domain-specific variation. Structured, data-driven tasks such as Weather, Speech-to-Text, and Flight Tracker achieve the highest cumulative similarity (total > 2.0 across metrics), reflecting clear functional boundaries that limit interpretive variance. In contrast, open-ended tasks like AI Face Generator, Sentiment Analysis, and Quotes yield the lowest combined similarity (< 1.3), showing that creative reasoning amplifies divergence. The composition plot indicates that AO contributes ≈ 31 % and τ ≈ 27 % of total similarity, underscoring that rank-order stability rivals retrieval overlap in shaping reliability.

The 3D landscapes (Figures 3a–b) show peaks for structured tasks and valleys for abstract ones. The smoother AO surface indicates greater stability in top-ranked APIs, aligning with cognitive patterns where LLMs agree on salient associations but diverge on marginal cases.

Volatility and consensus results (Figures 5–6) quantify stability across domains. Low variance in Weather and Speech-to-Text ($\sigma^2 \approx 3$–$4$) reflects deterministic reasoning, while high variance in Health and Sentiment ($\sigma^2 > 7$) shows sensitivity to sampling and phrasing. The Kemeny–Young distance follows this trend: volatility above 7 yields distances > 0.9, confirming that dispersion weakens agreement. These measures show domain structure as the key factor in inter-LLM stability—models align on objective tasks but diverge on semantic or affective ones.

Beyond metrics, the findings guide multi-agent coordination. Reliable tool use—e.g., in planning or reasoning—requires emphasizing stable clusters (Claude–DeepSeek–ChatGPT) and de-emphasizing volatile ones. Metrics like AO and Kemeny distance support this weighting, while high disagreement reveals overfitting or bias (e.g., ChatGPT–Mistral). Improving consistency demands aligned data, unified objectives, and systematic reliability benchmarking.

Overall, the study demonstrates that inter-LLM divergence is structured rather than random. Similarities cluster around well-defined tasks, while discrepancies arise systematically in ambiguous reasoning contexts. Improving cross-model consistency will therefore require not only better-aligned training data but also harmonized ranking objectives and calibration strategies—reinforcing the importance of systematic benchmarking for reliability-aware orchestration in multi-agent systems.

**Safety Implications for Multi-Agent Systems**

While this study focuses on quantifying agreement and divergence across LLMs, the findings have direct implications for AI safety. In multi-agent systems, coordinated behavior is often assumed when agents retrieve similar tools or actions. Our results show that this assumption is fragile: models may converge on a shared action space while internally disagreeing on priority and execution order.

Such latent divergence represents a class of failure modes that are difficult to detect using conventional single-agent or output-only evaluations. In safety-critical contexts, these inconsistencies may enable unintended coordination, brittle decision cascades, or strategically divergent behavior that only emerges through interaction.

By providing quantitative measures of inter-agent disagreement, volatility, and consensus distance, this framework can serve as a pre-deployment diagnostic tool. It enables practitioners and safety researchers to stress-test agent systems for hidden coordination failures prior to real-world execution, complementing existing interpretability and red-teaming approaches.

## Acknowledgements

This research work was supported in part by EA Funds (Long-Term Future Fund).

## Conclusion

This study introduced a unified framework for analyzing inter-LLM divergence in API discovery and ranking across fifteen tasks and five model families. Results show moderate overall agreement (AO ≈ 0.50, $\tau$ ≈ 0.45) but strong domain effects—structured tasks remain consistent, while open-ended ones diverge. Combining overlap, correlation, and consensus metrics reveals that reasoning stability decreases with abstraction and task complexity. These insights guide domain-adaptive orchestration, emphasizing trust in high-correlation clusters ($\tau > 0.6$) and verification in volatile domains (e.g., Sentiment Analysis).

Viewed through a safety lens, inter-LLM divergence provides an observable signal of latent coordination risk, underscoring the importance of benchmarked diagnostics for identifying misalignment and instability in autonomous agent systems before deployment. Future work will extend this framework to adversarial and paraphrased communication settings to evaluate whether divergence patterns persist, amplify, or collapse under perturbations—an important step toward detecting steganographic or strategically hidden coordination among agents.